\documentclass[11pt]{article}
\usepackage{hyperref}
\pdfoutput=1
\begin{document}
\title{Shock accelerated vortex ring}
\author{N. Haehn, C. Weber, J. Oakley, M. Anderson, D. Rothamer, R. Bonazza \\
\\\vspace{6pt} Dept. Engineering Physics
 \\ University of Wisconsin - Madison, Madison, WI 53706, USA}
\maketitle
\begin{abstract}
The interaction of a shock wave with a spherical density
inhomogeneity leads to the development of a vortex ring through the
impulsive deposition of baroclinic vorticity. The present fluid
dynamics videos display this phenomenon and were experimentally
investigated at the Wisconsin Shock Tube Laboratory's (WiSTL) 9.2 m,
downward firing shock tube. The tube has a square internal
cross-section (0.25 x 0.25 m$^2$) with multiple fused silica windows
for optical access. The spherical soap bubble is generated by means
of a pneumatically retracted injector and released into free-fall
200 ms prior to initial shock acceleration. The downward moving,
\emph{M} = 2.07 shock wave impulsively accelerates the bubble and
reflects off the tube end wall. The reflected shock wave
re-accelerates the bubble (reshock), which has now developed into a
vortex ring, depositing additional vorticity. In the absence of any
flow disturbances, the flow behind the reflected shock wave is
stationary. As a result, any observed motion of the vortex ring is
due to circulation. The shocked vortex ring is imaged at 12,500 fps
with planar Mie scattering.
\end{abstract}
\section{Introduction}
Shock-accelerated flows are important to many physical processes
encompassing a whole spectrum of length scales. When the large
pressure gradients inherent to shock waves encounter density
gradients resulting from inhomogeneities within the flow, vorticity
is deposited through the baroclinic mechanism ($\nabla \rho \times
\nabla p \neq 0$). This vorticity deposition magnifies any
perturbations within the density field and leads to the
Richtmyer-Meshkov Instability (RMI) \cite{richtmyer1954A}.

For Shock-Bubble-Interactions (SBI), the density inhomogeneity is
created by a spherical soap bubble. Significant work on SBI has been
performed by \cite{ranjan05,ranjan07}. The present study emphasizes
the interaction of a shock wave and a vortex ring generated during
the initial shock acceleration of the bubble. Planar Mie scattering
was used to image the shocked vortex ring at 12,500 fps. The shock
wave reflects off the tube end wall and encounters the vortex ring
formed from the first shock acceleration. Because the flow behind
the reflected shock wave is relatively stationary, imaging can take
place within a single 12 $\times$ 12 cm$^2$ fused silica window for
3-5 ms. The motion of the vortex ring is a direct result of the
circulation deposited upon the first and second shock accelerations.

The videos give an overview of the experimental setup, the
facilities at the Wisconsin Shock Tube Laboratory (WiSTL), the
method of generating and releasing a soap bubble, and the high speed
planar imaging of the shock accelerated bubble and vortex ring. The
videos can be found at either of the following links:

\begin{itemize}

    \item \href{http://ecommons.library.cornell.edu/bitstream/1813/14079/3/Gallery_Haehn_LowRes_v2.mpg}{Video
1 (Low Resolution)}
    \item \href{http://ecommons.library.cornell.edu/bitstream/1813/14079/2/Gallery_Haehn_v2.mpg}{Video
2 (High Resolution)}.

\end{itemize}

\bibliographystyle{unsrt}
\bibliography{biblio_database}

\begin{thebibliography}{1}

\bibitem{richtmyer1954A}
R.~D. Richtmyer.
\newblock Taylor instability in shock acceleration of compressible fluids.
\newblock {\em Physica D: Nonlinear Phenomena}, 12:1--3, 1984.

\bibitem{ranjan05}
D.~Ranjan, M.~H. Anderson, J.~G. Oakley, and R.~Bonazza.
\newblock Experimental investigation of a strongly shocked gas bubble.
\newblock {\em Physical Review Letters}, 94, 2005.

\bibitem{ranjan07}
D.~Ranjan, J.~Niederhaus, B.~Motl, M.~Anderson, J.~Oakley, and R.~Bonazza.
\newblock Experimental investigation of primary and secondary features in
  high-mach-number shock-bubble interaction.
\newblock {\em Physical Review Letters}, 98, 2007.

\end{thebibliography}

\end{document}